\documentstyle[12pt]{article}
\begin{document}
\title{\bf Increased Crowding during Escape Panic and Suitable Strategy for its Avoidance}
\author{R.~V.~R.~Pandya \thanks{Email: {\sl rvrpturb@uprm.edu}} \\
Department of Mechanical Engineering \\University of Puerto Rico at
Mayaguez\\ Puerto Rico, PR 00681, USA}
\date{}
\maketitle
\begin{abstract}
Under panicky situation, human have tendency to rush toward a particular direction for escape. I show here that this tendency alone causes increase in crowding and which could eventually trigger jamming that is not preferable. Further, it is proposed that potential flow theory can be employed in finding suitable strategy for escape. 
\end{abstract}

No one would prefer to be present in unstable situation of over-crowding leading to jamming and stampedes when individuals try to escape the situation under panic. Unfortunately, panicky situations are not completely avoidable as they are caused by forces, e.g. fires, whose origins in space and time remain unpredictable. We could only hope that spaces used by human and their escape routes were well laid out to avoid any casualty if panicky situation arises. The designs of layouts and appropriate escape routes require knowledge and mathematical representation of individual and collective behaviors of human during the escape. Based on the knowledge, we could adopt mainly two kinds of design strategies. In the first strategy (referred to as $S1$ hereafter), spaces and escape routes can be designed compatible to the natural behavior of human and without implementing any ways to change or manipulate human behavior during escape. This kind of strategy is usually employed in practice (see Ref. \cite{HFV00} and references cited therein). In the second strategy (referred to as $S2$ hereafter), the design can incorporate certain ways to change human behavior and their motions during the escape. Here, first I show that the natural behavior of human would always lead to over-crowding in $S1$ strategy and then I describe $S2$ strategy suggesting potential flow theory based motions for individuals to avoid or minimize the over-crowding. 

I put myself, in a hypothetical situation, among the people which were dancing and were dispersed uniformly in a large hall. Suddenly music stopped, there was a panic, reason unknown, and individuals started rushing directly toward the entrance, also an only exit for the hall. In the beginning of panic, I was able to move freely toward the exit with velocity $v_i^0(t){\bf e}_i^0(t)$. But soon I realized that I had to be careful not to bump into someone and my behavior was implicitly controlled by surrounding individuals and crowd. I started to have difficulty moving directly toward the exit and continued moving with velocity ${\bf v}_i(t)$ where ever I could, but was always trying hard to reach the exit. I was not fortunate enough and soon afterward got caught in the jam situation when the crowd was almost stand still and pushing, each other desperately, to squeeze through their way to the exit. People did not remain gentle then and what happened afterward is left for imagination. 

Though I was exhausted, I did realize that I experienced and felt the mathematical equation suggested by Helbing {\it et al.} \cite{HFV00}
\begin{equation}
m_i\frac{d{\bf v}_i(t)}{dt}=m_i\frac{{v}_i^0(t){\bf e}_i^0(t)-{\bf v}_i(t)}{\tau_i}+\sum_{j(\ne i)}{\bf f}_{ij}+\sum_{w}{\bf f}_{iw}.\label{eq1}
\end{equation}  
Here $m_i$ is $i$th individual's mass (e.g. my mass), ${v}_i^0(t){\bf e}_i^0(t)$ is intended velocity of an individual toward the exit due to natural tendency of human, ${\bf v}_i(t)$ is actual velocity of an individual with a certain characteristic time $\tau_i$ and $\sum_{j(\ne i)}{\bf f}_{ij}$ is sum of anticipated forces to keep myself a safe distance (if possible) from other surrounding individuals $j$'s and contact forces during pushing. Also, $\sum_{w}{\bf f}_{iw}$ is anticipated force to keep myself away from the nearest walls, had I been caught between the crowd and walls of the hall. The Lagrangian Eq. (\ref{eq1}) is the most reasonable mathematical description, available so far, for individual's behavior under the influence of surrounding individuals' behavior and presence of walls in panicky situation. Helbing {\it et al.} \cite{HFV00} considered a few benchmarks to show the robustness and usefulness of this equation for designing escape strategy of type $S1$.

So for strategy $S1$, now I show that the behavior governed by Eq. (\ref{eq1}) would always result in over-crowding if every individual in the hall rushes {\emph{directly}} toward the exit. For the discussion, consider magnitude $v_i^0(t)$ of the intended velocity to be identical for all individuals, independent of time and equal to $v^0$. But the unit direction vector ${\bf e}_i^0(t)$ would be different depending on the position of the individual with respect to the exit. Also, consider floor of the hall in $x-y$ plane of a rectangular coordinate system $x-y$ and exit of the hall at the origin. Every individual's position can then be given by coordinates $(x,y)$ and thus the intended velocity, directly toward the origin, of any individual located at $(x,y)$ can be given by the Eulerian velocity field ${\bf v}^0(x,y,t)$, written as
\begin{equation}
{\bf v}^0(x,y,t)= v^0 \Bigl[-\frac{x}{\sqrt{x^2+y^2}}\hat{i}-\frac{y}{\sqrt{x^2+y^2}}\hat{j}\Bigr]\label{eq3}
\end{equation}            
where $\hat{i}$ and $\hat{j}$ are unit vector along the $x$ and $y$ axes, respectively. Using the usual recipe of Kinetic Theory approach, Eulerian equations for number density $N(x,y,t)$ of people and number weighted velocity field ${\bf V}(x,y,t)$ can be obtained by using Eq. (\ref{eq1}) and Lagrangian equation $\frac{d{\bf R}_i(t)}{dt}={\bf v}_i(t)$ for position vector ${\bf R}_i(t)$ of each individual. The governing equation for $N$ and ${\bf V}$ can be written as, respectively,
\begin{equation}
\frac{D}{Dt}N\equiv [\frac{\partial }{\partial t}+{\bf V}\cdot {\bf \nabla} ]N= -N div {\bf V}, \label{num1}
\end{equation}
\begin{equation}
{\bf V}(x,y,t)= {\bf v}^0(x,y,t)-\tau_i\frac{D}{Dt}{\bf V}+{\bf V}^c, \label{eq4}
\end{equation} 
where ${\bf v}^0(x,y,t)$ is any intended velocity field (e.g. as given by Eq. \ref{eq3}) and  ${\bf V}^c$ accounts for the velocity due to forces ${\bf f}_{ij}$ and ${\bf f}_{iw}$. While writing Eq. (\ref{eq4}), time scale $\tau_i$ for all individuals is assumed to be identical. The three terms on the right hand side (rhs) of Eq. (\ref{eq4}) can be considered as zeroth order, first order and second order velocity terms, respectively, based on the order in which corresponding events causing these different velocities occur.  

In the beginning of the panicky situation ({\it Stage I}) when I was able to move freely, ${\bf V}^c={\bf 0}$ and I immediately gained intended velocity, i.e.  
\begin{equation}
{\bf V}(x,y,t)\cong {\bf v}^0(x,y,t), \label{s1}
\end{equation}
as suggested by Eq. (\ref{eq4}). In the beginning of {\it Stage II}, when I was avoiding collision with other individuals and was adjusting myself with small response or characteristic time scale $\tau_i$, ${\bf V}^c$ was negligible. For this stage, Eq. (\ref{eq4}) along with perturbation expansion around ${\bf v}^0(x,y,t)$ suggest
\begin{equation}
{\bf V}(x,y,t) \cong{\bf v}^0(x,y,t)-\tau_i [\frac{\partial }{\partial t}+{\bf v}^0\cdot {\bf \nabla} ]{\bf v}^0. \label{s2}
\end{equation}
These velocity fields of initial two stages are sufficient to show increase in over-crowding or in number density. 

Now during {\it Stage I}, Eqs. (\ref{eq3}) and (\ref{s1}) provide $div {\bf V}=-{v^0}/{\sqrt{x^2+y^2}}$ and its substitution into Eq. (\ref{num1}) yields 
\begin{equation}
DN/Dt=v^0N/\sqrt{x^2+y^2} > 0,
\end{equation}
suggesting increase in number density $N$ in time. This increase continues in time during the beginning of the {\it Stage II}. It should be noted that an additional $\tau_i$ containing term on the rhs of Eq. (\ref{s2}) can not reduce or increase the rate of increase, as $div[(\frac{\partial }{\partial t}+{\bf v}^0\cdot {\bf \nabla} ){\bf v}^0]=0$ when Eq. (\ref{eq3}) is used for ${\bf v}^0$.
This exhibits clearly that individuals' intentions to rush directly toward the exit is a main cause for increase in time of the number density of people i.e. over-crowding. 

Now I describe strategy $S2$. The cause of over-crowding is due to the compressible nature of the velocity field ${\bf V}$, i.e. non-zero value for $div {\bf V}$. The intended velocity field ${\bf v}^0$ arising from human's natural tendency is compressible too. So in order to avoid over-crowding and subsequent triggered events, the first required condition is that velocity field ${\bf V}$ should be incompressible i.e. $div {\bf V}=0$. Further, during escape I would not prefer to be moving in a circle again and again inside the hall. Thus, the second required condition of zero circulation suggests $curl {\bf V}=0$. These two conditions suggest the velocity field to be identical to that of inviscid, incompressible, irrotational fluid flow or potential flow, written in terms of a velocity potential function $\phi$ as
\begin{equation}
{\bf V}= grad\, \phi, \,\, \nabla^2 \phi =0.       
\end{equation}
Thus, if individuals move along the streamlines of appropriate potential flow with suitable speed, triggering of over-crowding and subsequent disastrous events can be altogether avoided. There is a possibility that a few individuals may not move exactly as required and perturbation in their behaviors arise due to finite response time to follow streamlines. The strategy $S2$ can then be further refined by using strategy $S1$. To do that, potential velocity field can be considered as intended velocity ${\bf v}^0$ in the framework of $S1$. The over-crowding will be caused at first order by the compressible nature of the second term on the rhs of Eq. (\ref{s2}). So for potential velocity field ${\bf v}^0$, divergence of Eq. (\ref{s2}) yields
\begin{equation}
div {\bf V}=-\tau_i\Bigl[\Bigl(\frac{\partial u }{\partial x} \Bigr)^2 +\Bigl(\frac{\partial v}{\partial y} \Bigr)^2+\frac{1}{2}\Bigl(\frac{\partial u }{\partial y}+\frac{\partial v}{\partial x} \Bigr)^2\Bigr] \,\, <0, \label{sr}
\end{equation}
where each term on the rhs is proportional to square of different strain rate. The Eqs. (\ref{sr}) and (\ref{num1}) suggest that over-crowding is likely to occur in the region of high strain rate of potential velocity field. Though it is not possible to have these strain rates zero everywhere, spaces and escape routes should be designed in a manner to minimize, as far as possible, areas of high strain rates in the potential velocity field. In addition, the quick response of human, i.e. small value for $\tau_i$, would be an added advantage to reduce the over-crowding.   

By now, we know suitable strategy. Now, important questions arise: How to implement $S2$ strategy in reality?  How to manipulate individuals' behaviors for their own advantage during escape so they quickly respond and move in, already established, streamlines pattern in panicky situations? I leave these difficult tasks for my fellow engineers to come up with clever ways, including and other than, marking streamlines of appropriate potential flow on the floor, using moving/flashing lights for speed at every location to guide individuals during escape.

\end{document}